\documentclass[aps,twocolumn,psfig,citesort]{revtex4}
\usepackage{epsfig}

\begin{document}

\title{Simulation study of Non-ergodicity Transitions: Gelation in Colloidal Systems 
with Short Range Attractions }
\author{Antonio M. Puertas}
\affiliation{Department of Physics, University of Almeria, 04120 Almeria, Spain}
\author{Matthias Fuchs}
\affiliation{Institut Charles Sadron, 67083 Strasbourg Cedex, France}
\author{Michael E. Cates}
\affiliation{School of Physics, The University of
Edinburgh, JCMB Kings Buildings, Edinburgh EH9 3JZ, UK}
\date{\today}

\begin{abstract}
Computer simulations were used to study the gel transition occurring in 
colloidal systems with short range attractions. A colloid-polymer 
mixture was modelled and the results were compared with mode coupling theory 
expectations and with the results for other systems (hard spheres and Lennard 
Jones). The self-intermediate scattering function and the mean squared 
displacement were used as the main dynamical quantities. Two different colloid
packing fractions have been studied. For the lower packing fraction, $\alpha$-scaling holds and
the wave-vector analysis of the correlation function shows that gelation is 
a regular non-ergodicity transition within MCT. The leading mechanism for 
this novel non-ergodicity transition is identified as bond
formation caused by the short range attraction. The time scale and diffusion 
coefficient also show qualitatively the expected behaviour, although different exponents 
are found for the power-law divergences of these two quantities. The non-Gaussian 
parameter was also studied and very large correction to Gaussian behaviour 
found. The system with higher colloid packing fraction shows indications of a nearby high-order singularity, causing $\alpha$-scaling to fail, but the general 
expectations for non-ergodicity transitions still hold. 
\end{abstract}
\maketitle

\section{Introduction}

Colloidal suspensions are often referred to as model systems for studying 
fundamental problems in condensed matter physics \cite{cates00}. Most of the 
properties of colloidal systems are similar to those of simple
liquids, except for the 
difference in the time scales involved in the processes in liquids or 
colloids, making the latter more useful in the study of some basic questions. 
Moreover, the interaction forces between particles in a colloidal system are 
easily tailored (e.g. by adding salt or polymer). However, there are some 
features found only in colloids, such as aggregation or gelation, which 
makes the study of these systems even more fascinating.

Gel formation, or gelation, is found in systems with strong short-range 
attractions, and is a universal phenomenon observed experimentally in many 
different systems, ranging from colloid-polymer mixtures \cite{poon99,dhont95}
to charged systems \cite{cipelletti00}, or to globular protein systems
\cite{muschol97}. Gelation is the formation of a percolating network 
(typically fractal) of dense and more dilute regions of
particles with voids which coarsen up to a certain size and freeze when the 
gel is formed. This process is observed in the structure factor as a low-$q$ 
scattering peak which moves to lower $q$, increasing its height, and then 
arrests \cite{carpineti92,poon97,segre01}. Description of this phenomenon has 
been attempted with percolation theories, theories of phase separation for 
states inside the liquid-gas binodal (which is meta-stable with respect to 
fluid-solid coexistence for short interaction ranges) or in terms of a glass 
transition of cluster of particles \cite{segre01,kroy03}.

Recently, acknowledging its non-equilibrium character, gelation has been 
interpreted using the formalism of mode coupling theory, MCT, for 
non-ergodicity transitions \cite{bergenholtz99,bergenholtz99a,bergenholtz00}.
This approach views the gel as particles trapped by a network of bonds 
which hinders the particle motion, resulting in a non-ergodic state. Thus,
gelation is caused by formation of long lived bonds, whose collective arrest is described as a a normal non-ergodicity transition. (This is distinct from many earlier approaches whereby the bonds were assumed to form irreversibly from the outset.)
In the present simulation study, we want to test this suggestion critically,
thereby establishing the existence or otherwise of a non-ergodicity transition
corresponding to bonding network formation.

Also present in colloidal systems is the equivalent of the usual glass 
transition in simple liquids,
which occurs at high densities, and is driven by steric imprisonment. This
transition has been studied experimentally and compared to MCT thoroughly
\cite{megen98,megen01,Beck99,Bartsch02}. When two different non-ergodicity 
transitions are observed in a system, MCT predicts a high order singularity 
in the region where the driving mechanisms for both transitions are present
\cite{Goetze88,gotze89,gotze02}. Therefore, a higher order transition is 
expected at high attraction strength and high density in colloidal systems 
with attractive interactions \cite{Fabbian99,bergenholtz99,dawson01}. 

Computer simulations have been used to test the expectations from MCT in 
many different systems, such as a Lennard-Jones liquid 
\cite{kob95,kob95a,gleim98,gleim00}, water \cite{sciortino96,sciortino97}, 
strong glass formers \cite{caprion00,sciortino00,horbach01} and polymers 
\cite{Bennemann98,Bennemann99,Bennemann99d,Aichele01}. The tests have shown 
that the predictions from MCT are correct, not only qualitatively but also,
in part, quantitatively \cite{nauroth97,sciortino00}. However, they have also 
pointed out some differences, especially in the spatial correlations of 
particle mobility \cite{donati99,donati99a,doliwa98,doliwa00}. In none of 
these simulated systems, however, did gelation occur, presumably because the
attractions were not short-ranged enough.

In this work, we have used molecular dynamics simulations to study the 
properties of the gel transition, and compared them with the predictions 
from MCT. (This was initiated in \cite{puertas02} where some further results 
may be found.) We take the numerous universal predictions of the theory to test
the scenario qualitatively. Comparing with quantitative predictions
available for systems of hard spheres \cite{fuchs98,fuchs92}, spheres
with short range attractions
\cite{bergenholtz99,bergenholtz99a,bergenholtz00,dawson01}, and the
mentioned simulation studies, we identify the novel mechanism driving the
non-ergodicity transition which is the cause of gelation for moderately
dense suspensions. Molecular dynamics were used instead of Brownian dynamics
because the choice of microscopic dynamics does not affect the 
relaxational dynamics of a
system  close to a non-ergodicity transition \cite{gleim98}. By means of the
Asakura-Oosawa interaction potential \cite{asakura54}, we simulate the 
behaviour of a colloid-polymer mixture, which is a well-understood system
\cite{gast83,gast86,dijkstra98,dijkstra99}. For short interaction ranges, 
this system exhibits a fluid-crystal transition, at intermediate densities 
and increasing attraction strength, with a liquid-gas transition meta-stable 
to the fluid-crystal one. In our simulations, the system was modified to 
prevent both of these phase transitions from occurring, in order to be able 
to study the transition from the fluid to the non-equilibrium states.

The paper is organized as follows: Section II describes some results from MCT 
which will be used in the subsequent analysis of the simulation results. In 
section III the simulation method is presented and the details are given. 
Section IV, deals with the results and is divided into four subsections
studying {\sl i}) the correlation function, {\sl ii}) the time scale and the
diffusion coefficient, {\sl iii}) the mean squared displacement and {\sl iv})
a higher colloid concentration. Finally, in section V, we present 
the conclusions of this work.

\section{Mode Coupling Theory}

In this section we will present the most important MCT results on 
non-ergodicity transitions. MCT attempts a description of the
density correlator and its self part, in terms of a fluctuating-force
correlator \cite{gotze91,gotze92}. In this paper, only the self part of 
the density correlator will be studied, defined as:  

\begin{equation}
\Phi_q^s(t)\:=\:\langle \exp\left\{i {\bf q}
\left({\bf r}_j(t)-{\bf r}_j(0)\right) \right\} \rangle
\end{equation}

\noindent where the brackets denote average over particle $j$ and time
origin, and ${\bf q}$ is the wave-vector. The equation of motion of
$\Phi_q^s$ in Brownian (coarse grained) dynamics, is given by: 

\begin{equation}
\tau_q \partial_t \Phi_q^s(t)+\Phi_q^s(t)+\int_0^tm_q(t-t')
\partial_{t'}\Phi_q^s(t') dt'\:=\:0 
\end{equation}

\noindent where $\tau_q$ is a single particle diffusive time scale and
$m_q(t)$ is a mode coupling kernel which describes the cage effect
\cite{hansen86}. Within MCT, glass states are given by non-zero solutions
of this equation for the long time limit of $\Phi_q^s(t\to\infty)=f_q^s$, 
the so-called {\sl non-ergodicity parameter}. It describes the
glass structure and may also be called {\sl Lamb-M\"ossbauer factor}.
The glass transition is marked by a (generally) discontinuous transition
from the unique trivial solution in the liquid, $f_q^s=0$, to multiple
solutions in the glass, $f_q^s>0$, where only the highest solution is 
physical. Glass transitions can be classified according to the number,
$l-1$, of non-trivial solutions merging with the highest one, and the type
of transition is noted as $A_l$.  

For liquid states close to the glass, a two step decay is observed for the
correlator; the plateau is at $f_q^s$ and signals the proximity of the
glass transition. Around this plateau, $\Phi_q^s$ shows some universal
properties, depending on the type of transition. For the most common type
of transition, $A_2$, the decay to the plateau, and that from the plateau, can
both be expressed as power law expansions. In particular, the decay from the
plateau is given by: 

\begin{equation}
\Phi_q^s(t)\:=\:f_q^s-h_q^{(1)}\left(t/\tau\right)^b+h_q^{(2)} 
\left(t/\tau\right)^{2b}+O(\left(t/\tau\right)^{3b})
\label{a-decay}\end{equation}

\noindent with $h_q^{(1)}$ and $h_q^{(2)}$ amplitudes and $\tau$ the final
or $\alpha$-relaxation time scale. $b$ is known as the von Schweidler
exponent, and depends on the details of the interaction potential.
Expression (\ref{a-decay}) implies time scaling for the decay from the
plateau, called $\alpha$-decay, for different states close to the glass 
transition.
The time scale, $\tau$, diverges as the glass transition is approached
according to a power law, with an exponent $\gamma$, which can be related to
the von Schweidler exponent: $\tau \sim |\sigma| ^{-\gamma}$, with $\sigma$
the distance to the transition \cite{gotze91,gotze92}. On the other hand,
the wave-vector dependence of the non-ergodicity parameter and amplitudes
gives some non-universal properties of the transition, providing information
about the mechanism causing the non-ergodicity transition.

For high order singularities, the fluid states close by show again a two
step decay in the correlation function, but the decays to and from the 
plateau are no longer power law expansions. Instead, logarithmic laws are 
obtained \cite{gotze89,gotze02}. A salient feature is that a logarithmic decay 
around the plateau is predicted:

\begin{equation}
\Phi_q^s(t)\:=\:f_q^{s A}-C_q \log \left( t/t_1 \right)
\label{log-decay} \end{equation}

\noindent where $f_q^{s A}$ is the non-ergodicity parameter of the high 
order singularity, $C_q$ is an amplitude and $t_1$ is a time scale (the time 
when the correlator lies on the plateau). 

The mean squared displacement (MSD) can be studied instead of the correlation 
function, obtaining a similar two step behaviour. Similar asymptotic laws 
to describe the decay to and from the plateau can be derived, and the 
parameters and exponents can be related to those of the correlation function 
\cite{fuchs98}. The value of the plateau in the MSD defines the {\sl
localization length} and is a measure of the size of the cage. However, it
should be noticed that the cage, as formed by other particles, is constantly
restructuring cooperatively. Only when the particles have broken free of
their cages, diffusive motion is observed, with a self-diffusion
coefficient, $D_s$, that tends to zero as the glass transition is approached
as $D_s\sim |\sigma|^{\gamma}$ for the usual $A_2$ transitions. 

Two different non-ergodicity transitions have been found in colloidal systems 
with a short range attraction \cite{bergenholtz99,dawson01}: a steric 
hindrance driven glass transition and an attraction driven gel transition 
\cite{bergenholtz99,bergenholtz99a,bergenholtz00}. While the first is found 
at high densities and is qualitatively similar to the glass transition in
the hard sphere system (HSS) or Lennard-Jones system (LJS), the gel
transition occurs at high attraction strength for all volume fractions.
Different properties for these two transitions are predicted, the main
difference arising from the driving mechanism: the localization length is
shorter in the gel than in the glass, resulting in higher non-ergodicity
parameters. Also, a smaller von Schweidler exponent for the gel than for the
glass is expected, implying a higher value of $\gamma$, i.e. the transition
as observed by $\tau_q$ or $D_s$ is more abrupt.

The actual shape of the non-ergodicity transition line depends on the
details of the interaction potential, although some general features can be
found. From lower to higher interaction strength, the glass line, is slanted
to higher concentrations, showing that a weak attraction {\sl fluidizes} the
glass. However, at even higher interaction strengths, the gel transition
occurs at lower colloid density the higher the attraction strength. As a
result, a re-entrance transition is obtained at high colloid volume
fractions. The line may be wedge-shaped or curved in this region, depending
on the range of the interaction. If the line is wedge-shaped a high-order
transition (generically $A_3$) is present near the corner, whereas none exists 
if the line is smoothly continuous. An $A_4$ singularity appears right at the 
vanishing of the $A_3$ point when the line first becomes smooth \cite{dawson01}.

\section{Simulation details}

Equilibrium molecular dynamics simulations mimicking a colloid-polymer mixture 
were performed for a system composed of 1000 soft-core polydisperse colloidal 
particles. The core-core interaction between particles was modeled by:

\begin{equation}
V_{sc}(r)\:=\:k_BT \left(\frac{r}{a_{12}}\right)^{-36}
\end{equation}

\noindent where $a_{12}=a_1+a_2$, with $a_1$ and $a_2$ are the radii of the 
interacting particles. A flat distribution of radii with a width of 
$\delta=0.1 a$, where $a$ is the mean radius, was used. The exponent in 
$V_{sc}$ was selected high enough to avoid 
problems related to the softness of the potential \cite{melrose92}. The 
polymer induces an attractive depletion interaction between the colloidal
particles, which was modeled by the Asakura-Oosawa interaction potential
\cite{asakura54, dijkstra99}. The extension of this potential to take
polydispersity into account reads \cite{mendez00}:  

\[ V_{AO}(r) \:=\: -k_BT \phi_p \left\{\left[\left(\bar{\eta}+1\right)^3
-\frac{3r}{4\xi} \left(\bar{\eta}+1\right)^2+\frac{r^3}{16\xi^3}\right]+
\right.\]
\begin{equation}\label{pot}
\left.+\frac{3\xi}{4r} \left(\eta_1-\eta_2\right)^2 \left[\left(\bar{\eta}+1
\right) -\frac{r}{2\xi} \right]^2\right\}
\end{equation}

\noindent for $r \leq 2(a_{12}+\xi)$ and 0 for larger distances. Here, 
$\eta_i=a_i/\xi$; $\bar{\eta}=(\eta_1+\eta_2)/2$, and $\phi_p$ is the 
volume fraction of the polymer. Note that the range of the potential is
given by the polymer size, $\xi$, and its strength by $\phi_p$. This potential 
was modified around $r=a_{12}$, to ensure that the minimum of the total 
potential ($V_{sc}+V_{A0}$) occurs at this point: for $r \leq 2a_{12}+\xi/5$ 
a parabolic form, which connects analytically to $V_{AO}$ at $2a_{12}+\xi/5$
and has a minimum in $2a_{12}$, was used. In our simulations, the range of
the interaction, $2\xi$, was set to $0.2 a$, which would
correspond to polymers with $R_g/a=0.1$ where $R_g$ is the radius of gyration.

\begin{figure}
\psfig{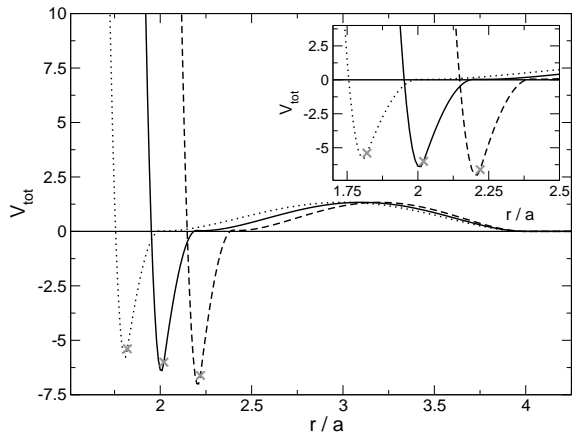}
\caption {\label{potential}
Total pair interaction potential $V_{\rm tot}$ as function of the
radial distance $r=|{\bf r}_1-{\bf r}_2|$ for three different particle
pairs; a pair of 
particles with minimal radii $a_1=a_2=a-\delta$, one with average radii
$a_1=a_2=a$, and one with maximal  $a_1=a_2=a+\delta$ (from left to
right). The inset shows the enlarged  region of the attractive minimum.
Crosses mark where the parabolic minimum smoothly matches to
eq. (\protect\ref{pot}).}
\end{figure}

A long-range repulsive barrier was added to the interaction potential in order 
to prevent liquid-gas separation (as shown below). The barrier had a maximal 
height of $1 k_BT$, according to a fourth-order polynomial:

\begin{equation}
V_{bar}(r)\:=\:k_BT\left\{\left(\frac{r-r_1}{r_0-r_1}\right)^4-2\left(
\frac{r-r_1}{r_0-r_1}\right)^2+1\right\}
\end{equation}

\noindent for $r_0\leq r \leq r_1$ and zero otherwise. The limits of the 
barrier were set to $r_0=2(a_{12}+\xi)$, and $r_1=4a$, which was enough to 
prevent phase separation. The maximum height of the barrier equals the depth 
of the depletion interaction at contact for $\phi_p=0.0625$, much lower than
the values where the gel transition takes place. The resulting total
interaction potential, $V_{tot}=V_{sc}+V_{AO}+V_{bar}$, is analytical 
everywhere. It is shown in figure \ref{potential}, where in order to
indicate the spread induced by polydispersity, the potentials among three
different pairs with differing radii are plotted.

In our simulations, lengths were measured in units of the mean radius, $a$,
and time in units of $\sqrt{4a^2/3v^2}$, where the thermal velocity $v$ was
set to $\sqrt{4/3}$. Equations of motion were integrated using the
velocity-Verlet algorithm, in the canonical ensemble (constant NTV), to
mimick the colloidal dynamics. Every $n_t$ time steps, the velocity of the
particles was re-scaled to assure constant temperature. No effect of $n_t$ was
observed for well equilibrated samples. The time step was set to $0.0025$.
Equilibration of the systems was tested by monitoring the total energy, and
other order parameters (see below), and by measuring $\Phi_q^s(t)$ and the
MSD at different initial times. When the order parameters were constant and
the $\Phi_q^s(t)$ and MSD curves showed no dependency on the initial
time (ageing), the system was considered to be equilibrated.

The volume fraction of the colloidal
particles, $\phi_c=\frac{4}{3}\pi a^3 \left(1+\left(\frac{\delta}{a}\right)^2
\right) n_c$, with $n_c$ the colloid number density, and the polymer volume
fraction, $\phi_p$, were the control parameters used to identify the states in
the phase diagram.

\begin{figure}
\psfig{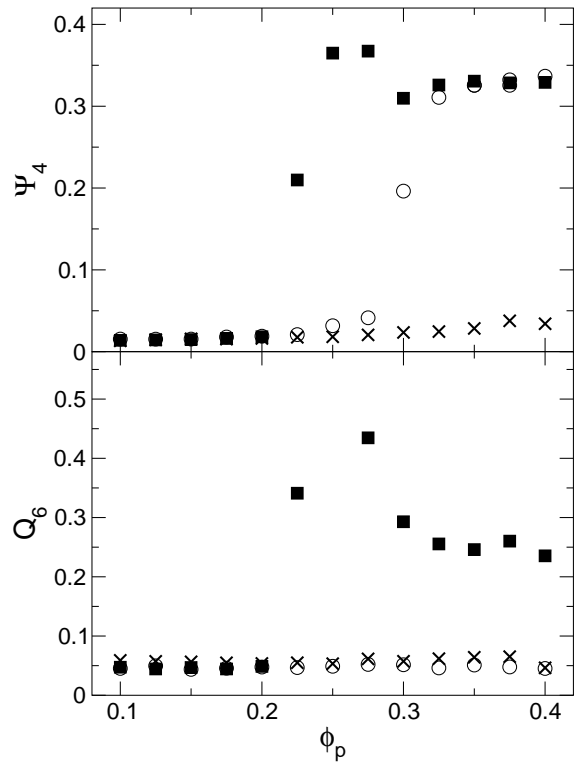}
\caption {\label{phase_diagram}
Demixing ($\Psi_4$) and orientational ($Q_6$) parameters for 
$\phi_c=0.40$ and increasing polymer fraction, $\phi_p$, for different 
systems: monodisperse without long-range barrier (squares), polydisperse 
without barrier (circles), and polydisperse with long-range barrier (crosses).}
\end{figure}

In order to explore the whole $\phi_p-\phi_c$ plane in search of the gel
transition, phase transitions which forbid access (in equilibrium) to 
important 
parts of the plane must be prevented. Several order parameters were used to 
identify different kinds of ordering in our system and to 
monitor whether unwanted liquid-gas or fluid-crystal transitions were taking 
place. First, the onset of phase separation involving states of different 
density can be detected by 
dividing the system into $n^3$ boxes and measuring the density in every box. 
The `demixing' order parameter is defined as the standard deviation of the 
distribution of densities:

\begin{equation}
\Psi_n\:=\:\sum_{k=0}^{n^3} \left( \rho_k-\bar{\rho} \right)^2
\end{equation}

\noindent where $\rho_k$ is the density of particles in box $k$, and 
$\bar{\rho}$ is the mean density. This parameter is close to zero for an 
homogeneous system, and increases if it demixes into phases of different 
density. In our case, $n$ has been set to $4$, implying $64$ boxes, and a 
box edge of about $5a$ (depending on $\phi_c$). On the other hand, the 
orientational order parameter, $Q_6$, as defined by Steinhardt et al. 
\cite{steinhardt83,frenkel96}, signals the presence of an ordered phase, and
is used to detect crystallization.

The phase diagram was probed using these parameters. In figure 
\ref{phase_diagram} the results are presented for a bare system (monodisperse 
and without the long-range barrier), a polydisperse system without the 
long-range barrier, and the final system with both polydispersity and
barrier. In this figure, the colloid volume fraction is constant,
$\phi_c=0.40$, and the polymer concentration varies; an isochore is studied. 
The sudden increase in both $\Psi_4$ and $Q_6$ occurring at 
$\phi_p=0.20$ for the bare system, signals the crystallization boundary, in 
accordance with Dijkstra et al. \cite{dijkstra99}. Because of the short range 
of the potential, this system has no liquid phase; i.e. the liquid-gas 
coexistence is meta-stable with respect to the crystal-gas transition.

\begin{figure}
\psfig{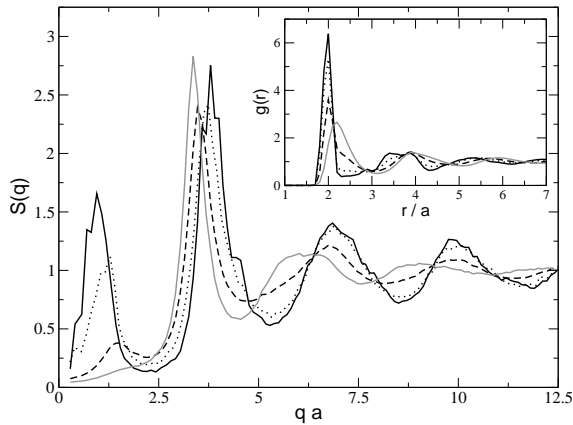}
\caption {\label{sdq}
Structure factors for different polymer fractions at $\phi_c=0.40$:
$\phi_p=0$ gray line, $\phi_p=0.2$ dashed line, $\phi_p=0.35$ dotted line
and $\phi_p=0.425$ solid black line. Note the low-q peak raise as $\phi_p$
is increased. Inset: Pair distribution function, $g(r)$ for the same
states.}
\end{figure}

When polydispersity is introduced in the system, crystallization is
prevented, as indicated by the constant trend of both parameters close to
$\phi_p=0.20$. However, as the system now does not crystallize, the
liquid-gas transition can be reached upon increasing the strength of the
interaction. This demixing is signalled by an increase in $\Psi_4$, not
involving local ordering. In order to avoid this separation, the long-range
barrier has been introduced in the interaction potential. The energy of a
dense phase is raised, and demixing is thus energetically unfavourable.
Figure \ref{phase_diagram} shows that liquid-gas separation is indeed
inhibited by the repulsive barrier. Instead, individual voids of finite size
are created in the system, causing a low-q peak in the structure factor,
$S(q)$, presented in Figure \ref{sdq}. There, $S(q)$ is shown for different
polymer fractions, ranging from no attraction ($\phi_p=0$) to the closest
state to the gel we have accessed ($\phi_p=0.425$). 

In the inset to Figure \ref{sdq}, the pair distribution function, $g(r)$, is
presented for the same states as the structure factor. The value at contact,
$r=2$, increases continuously as the atraction strength grows,
signalling increased local contact probabilities. This process will be shown 
to be responsible for the gel transition. In $S_q$ it becomes evident as an 
increase in the oscillations for large $q$.

The low-q peak in the structure factor resembles the low-angle peak observed
in light scattering experiments with colloidal gels
\cite{carpineti92,segre01}. However, whereas the peak in our system is an
equilibrium property, induced by the specific shape of the interaction
potential, the experimental peak has a non-equilibrium origin. 
We also checked for the possibility of microphase separation, which in some 
cases can be induced by a repulsive barrier \cite{sear99}. In our case, the 
small angle peak continuously increases with $\phi_p$, but stays finite and 
smaller than the neighboring peak as we approach the gel transition. We 
interpret this to indicate that we do not have microphase separation, and
we also observed no other signs of such ordering. Furthermore, since
the relevant wave vectors in the MCT calculation of the gel transition are
the high ones (around $2\pi/\xi$), the change in low-q region in the
structure factor is expected to have little effect on the gel transition.  

\section{Results and Discussion}

This system has been previously shown to undergo both the glass and gel 
transitions as stated by MCT. It also exhibits a logarithmic decay in the
correlation function at high colloid concentration, indicating a high-order
singularity in that region \cite{puertas02}. In this section we will 
discuss the properties of the gel transition, and compare them with MCT and
with those of the HSS and other systems, which are similar to the glass 
transition at high colloid concentration. We test for differences by
comparing quantitatively the non-universal features of the transition, which
will aid in the identification of the driving mechanism.

The gel line is predicted to extend to low packing fractions with the same
qualitative properties. In order to test these properties we have performed 
simulations at two different colloid concentrations, $\phi_c=0.40$ and 
$\phi_c=0.50$, where the gel line is far away from the percolation one. At 
high concentration, the higher order singularity is expected to affect the 
equilibrium states, disturbing some features of the gel transition.

\subsection{Self-Intermediate Scattering Function}

\begin{figure}
\psfig{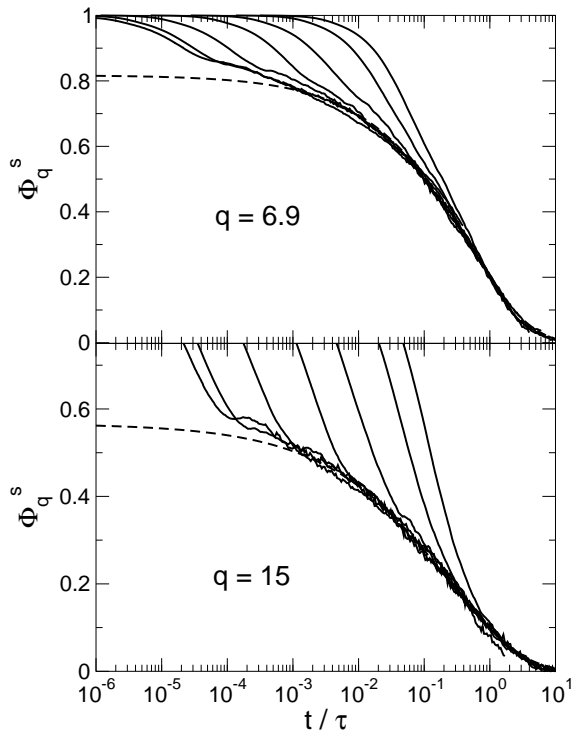}
\caption {\label{scaling}
Intermediate scattering function (self part), $\Phi_q^s$, vs. re-scaled 
time with the $\alpha$-time scale, $\tau_q$, for different states:
$\phi_c=0.40$ and $\phi_p=0.375,\,0.39,\,0.40,\,0.41,\,0.415,\,0.42$ 
and $0.425$ from right to left. Two different wave-vectors are 
studied: $q=6.9$ (upper panel) and $q=15$ (lower panel), with the
KWW fits (dashed line) included.}
\end{figure}

The scaling prediction for the $\alpha$-decay of states close to a 
non-ergodicity transition is tested in Figure \ref{scaling}, for constant 
colloid packing fraction, $\phi_c=0.40$. Two different
representative wave-vectors are 
presented in this Figure, $q=6.9$ and $q=15$. As observed at the glass 
transition in the HSS and many other different systems 
\cite{kob95,sciortino97,kob95a,horbach01,Aichele01,puertas02}, the 
$\alpha$-scaling property holds. In comparing these correlation functions
with those typical for the HSS or LJS, it is noticed that in Figure
\ref{scaling} the $\alpha$-decay of the correlators is more stretched,
implying a smaller von Schweidler exponent at the gel transition than at
the glass transition. Because of this stretching in the $\alpha$-decay, a
clear plateau is not observed, although a slowing down of more than four
decades is studied. Nevertheless, extrapolating the relaxation curves to
extract plateau values, much higher values are found than at the glass 
transition in the HSS or the LJS.

\begin{figure}
\psfig{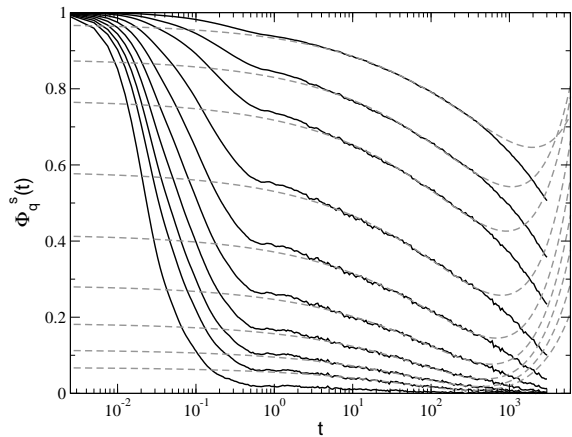}
\caption {\label{fsqt} Intermediate scattering function (self part), 
$\Phi_q^s$, for the state $\phi_c=0.40$ and $\phi_p=0.42$, for different 
wave-vectors; from top to botton:
$q=3.9,\,6.9,\,9.9,\,15,\,20,\,25,\,30,\,35,\,40,\,50$. The dashed lines are
fittings from eq. (\ref{a-decay}) up to second order, with the same von 
Schweidler exponent for all $q$.}
\end{figure}

We have analyzed the state $\phi_c=0.40$
and $\phi_p=0.42$ in more detail, which shows four decades of slowing 
down compared to the purely repulsive situation upon turning on the
attraction. Because scaling is observed in Figure \ref{scaling}, studying
only one state is enough to analyse the $\alpha$-decay of the correlation
function. The slowest state, $\phi_p=0.425$ was not chosen because it
strongly deviates from the expected behaviour of $\tau_q$ vs. $\phi_p$ (see
Figure \ref{tauq_phi} and discussion thereafter). The correlation functions
at different wave-vectors for state $\phi_c=0.40$ and $\phi_p=0.42$ are
presented in Figure \ref{fsqt}. The range of wave-vectors studied, where the
plateau height changes is much wider than the range for a similar change in 
$f_q$ at the glass transition of hard spheres or Lennard-Jones particles. This 
feature indicates that the relevant distances for the gel transition are much
shorter than for the usual glass transition. 

The correlation functions were measured until the average particle 
displacement was $5a$, which is one fourth of the box size ($21.95\,a$). 
Thus, extending this measurement to longer times in order to observe the
whole $\alpha$-decay at low $q$ is troublesome. If the 
diffusion coefficient diverges at the same rate as the $\alpha$-time scale 
(as predicted by MCT), this problem would not appear. Thus, we are also 
observing a discrepancy between both time scale divergences, that will be 
further discussed below.

\begin{figure}
\psfig{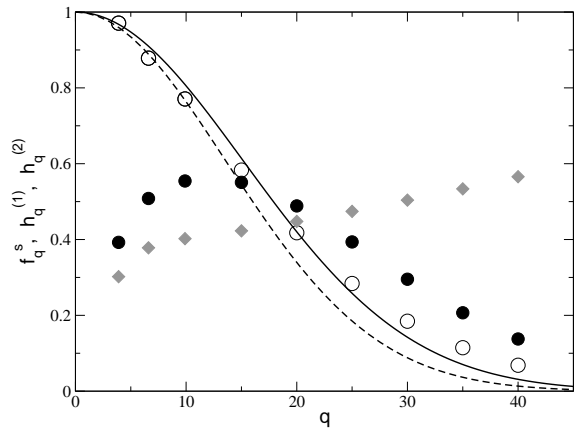}
\caption {\label{fq} Non-ergodicity parameter, $f_q^s$ (open circles), and 
first (filled circles) and second order (diamonds) amplitudes from the 
fittings in Figure \ref{fq} with $b=0.37$ for all  wave-vectors. The lines
give the Gaussian approximation from eq. (\protect\ref{gauss}) fitted to all 
wave-vectors (solid line) and the three lowest wave-vectors (dashed line).}
\end{figure}

The impossibility of observing a clear plateau, as mentioned above, makes it
more difficult to analyse the correlators, since $f_q^s$ cannot be fixed a
priori. Therefore, the $\alpha$-decay of the correlation functions has been
fitted using expression (\ref{a-decay}), with $f_q^s$, $h_q^{(1)}$ and
$h_q^{(2)}$ as fitting parameters. The von Schweidler exponent was also
fitted but was kept identical for different wave-vectors. It was found as
$b=0.37$, and the other results for the fitting parameters, are shown in
Figure \ref{fq}. The trends of these parameters are similar to that of the
glass transitions in both HSS and LJS, but over a wider q-range in the gel
case. This indicates that the localization length is quite different in the 
present system. The non-ergodicity parameter exhibits a bell shaped curve, 
whereas the first order amplitude describes a maximum. The latter
is determined from the fit up to a prefactor which depends on the choice of
$\tau$ in eq. (\ref{a-decay}). As an estimate we have used $\tau_q$ for
$q=9.9$ ($\tau_q$ is defined by $\Phi_q^s(\tau_q)=f_q/e$), which yields
values that are similar (in magnitude) to the HSS. The second amplitude
shows a monotonously increasing behaviour with $q$, in accordance with the
HSS, but it is always positive, unlike the HSS where it goes through zero at
the peak of $h_q^{(1)}$.  

The non-ergodicity parameter, $f_q^s$, can be approximated using the 
Gaussian expression:

\begin{equation}\label{gauss}
f_q^s\:\approx \: \exp \left\{ -q^2 r_l^2 /6\right\}
\end{equation}

\noindent where $r_l$ is the localization length. This approximation 
is known to be valid for low wave-vectors, and important deviations from the 
Gaussian behaviour are expected close to the glass transition. However, the 
value for the localization length obtained from fitting this curve (solid
line in Fig. \ref{fq}), can be used as an estimate of the one in the MSD.

The localization length so obtained is $r_l^2=0.0126 a^2$, much smaller than
for the HSS or the LJS, where $r_l$ is of the order of the Lindemann
distance. This feature shows that the process causing the non-ergodic
transition in our case has a typical distance much smaller than in the case
of glass transitions in the HSS or the LJS. This agrees with the observation
and discussion about the height of the plateaus, and of the different q-range
covered by $f_q^s$ in Figure \ref{fq}. Whereas the glass transition in the
HSS is driven by core-core repulsions, the gel transition is caused by the
short range attraction, therefore by {\sl bonds} between particles, (see
inset to Figure \ref{sdq}) whose
size is of the order of the interaction range. An interesting analogy has
been established between the mechanisms driving the formation of gels and
glasses, and the freezing transition \cite{foffi02}.  

The $\alpha$-decay of near-non-ergodic states can be also studied 
using the Kohlrausch-Williams-Watts (KWW) stretched exponential. The KWW 
expression is given by:

\begin{equation}
\Phi_q^{K}(t)\:=\:A_q \exp \left\{ - \left( 
\frac{t}{\tau_q^{K}}\right)^{\beta_q} \right\} \label{kww} 
\end{equation}

\noindent where $\beta_q$ is known as Kohlrausch exponent, which has been 
shown to coincide with the von Schweidler exponent at high wave-vectors 
\cite{fuchs94}. This expression has been fitted to very different systems, 
and describes the $\alpha$-decay down to zero. 
We have fitted this expression to the $\alpha$-decay in our system. However, 
since the correlators in Figure \ref{fsqt} do not show the complete 
$\alpha$-decay, we have fitted expresion (\ref{kww}) to the master curve, 
obtained from the $\alpha$-rescaling. Two of these fittings are presented in 
Figure \ref{scaling} by the dashed lines, showing that the KWW stretched
exponential describes well the $\alpha$-decay in this system.

\begin{figure}
\psfig{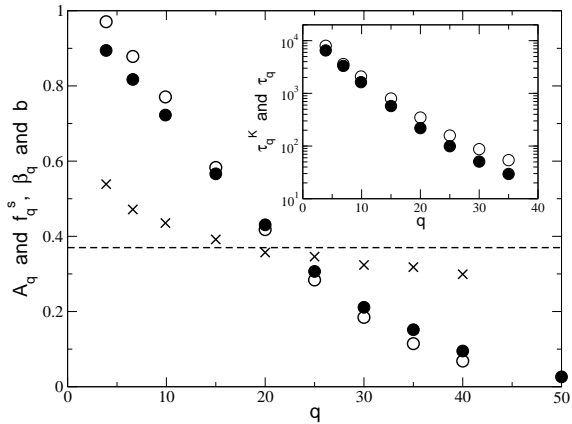}
\caption {\label{a_q}
Parameters used in the KWW fittings and comparison with the 
von Schweidler fitting. Main figure:  $A_q$ (closed circles) and $f_q^s$ 
(open circles), $\beta_q$ (crosses) and $b$ (horizontal dashed line). Inset: 
$\tau_q^K$ (closed circles) and $\tau_q$ (open circles).}
\end{figure}

The fitting parameters, $A_q$, $\beta_q$ and $\tau_q$, are presented in 
Figure \ref{a_q}, and compared with the corresponding parameters in the von 
Schweidler formalism. In such a way, $A_q$ is compared with the non-ergodicity
parameter, $\beta_q$ with the von Schweidler exponent, and the $\tau_q^K$ with
$\tau_q$. As expected, the height of the plateau can be determined equally 
well both by the KWW or von Schweidler analysis. The same holds for the time 
scales, $\tau_q^{KWW}$ and $\tau_q$. The Kohlraush exponent is expected to 
tend to $1$ at low wave-vectors, and to approach the value of the von 
Schweidler exponent at high $q$. The low-$q$ limit is explained because 
diffusion is the dominant process over long distances, whereas at short 
distances (comparable to the cage size) the dynamics is dominated by the 
cooperative local rearrangements. This behaviour is predicted from MCT 
\cite{fuchs94}, and has been observed in different systems, such as molecular 
glass formers \cite{Toelle98}, and in simulations of polymer melts 
\cite{Aichele01}, and of water \cite{Starr99}. In our case, the low-$q$ limit 
is not observed, but $\beta$ rises as the wave-vector decreases, indicating 
that the expected behavior may appear at lower $q$ below the small angle peak
in $S(q)$. At high wave-vector, the Kohlraush exponent crosses the von 
Schweidler value, but stays close to it. Although an exact agreement is not 
observed, we may conclude that the correct general trend is obtained.

\subsection{Time Scale and Diffusion Coefficient}

An important universal prediction of MCT is the existence of power law 
divergences for both the time scale, $\tau$, and the inverse of the self 
diffusion coefficient $D_s$, with the same exponent in both cases, $\gamma$:

\begin{equation}
\tau_q\sim \left(\phi_p^G-\phi_p \right)^{-\gamma} \hspace{0.5cm} \mbox{and} 
\hspace{0.5cm} D_s\sim\left(\phi_p^G-\phi_p\right)^{\gamma} \label{power-law}
\end{equation}

\noindent where $\phi_p^G$ is the polymer volume fraction where the gel
transition occurs. The relation between exponent $\gamma$ and the von
Schweidler exponent, $b$, is also universally established by MCT
\cite{gotze92}. 

Testing of the power law divergence (and measuring of $\gamma$) is usually 
carried out plotting $\tau_q$ as a function of $\phi_p^G-\phi_p$ for different 
values of $\phi_p^G$, looking for a straight line. This method is cumbersome, 
even more as deviations from it are expected for states close to the 
transition, and precise values for $\gamma$ and $\phi_p^G$ cannot be given. 
To avoid this difficulty, we have calculated $\gamma$ from $b$, as given by 
MCT, and with this particular value of the exponent looked for the power law 
divergence. In such a way, we are testing the {\sl compatibility} of MCT 
predictions with our data.

\begin{figure}
\psfig{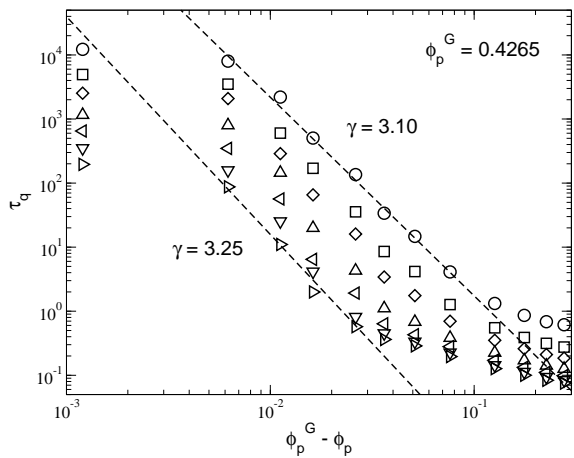}
\caption {\label{tauq_phi}
Wave-vector dependent time scale, $\tau_q$, vs. $\phi_p^G-\phi_p$ for the
isochore $\phi_c=0.40$ for different wave-vectors; symbols from top to
botton correspond to $q=3.9,\,6.9,\,9.9,\,15,\,20,\,25,\,30$. The lines are
power law fittings to $q=3.9$ and $q=30$. In all cases $\gamma$ keeps close
to this two values.} 
\end{figure}

Figure \ref{tauq_phi} shows the wave-vector dependent time scale $\tau_q$ vs. 
$\phi_p^G-\phi_p$ for different wave-vectors. For every wave-vector, 
$\tau_q^{1/\gamma}$ was 
extrapolated to zero, yielding a value for the polymer fraction at the gel 
transition, $\phi_p^{G,q}$. The final value of $\phi_p^G$, used in Figure 
\ref{tauq_phi}, was calculated as the average value for all wave-vectors 
studied. The linear trends in Figure \ref{tauq_phi} for  $\phi_p^G-\phi_p>5 
\cdot 10^{-3}$, shows the power-law behaviour predicted by MCT, with exponent 
$\gamma=3.1$ and $\phi_p^G=0.4265$. The 
closest state to the gel transition, $\phi_p=0.425$ deviates from the 
power-law behaviour observed for lower polymer fractions. Similar deviations 
have been observed in the HSS and LJS and can tentatively be attributed to 
thermally activated processes (or hopping events) \cite{note1}.

As shown in eq. (\ref{power-law}), MCT predicts a
power-law for the self diffusion coefficient, $D_s$, with the same  
exponent as the divergence of the time scale. Simulations on HSS 
and LJS have shown that a power law divergence is indeed obtained, but with
a different exponent than in the case of $\tau_q$. Using the same procedure 
as described above (calculating $\gamma$ from $b$ and extrapolating 
$D_s^{-1/\gamma}$ to obtain $\phi_p^G$) 
yields a value for $\phi_p^G=0.4519$, with the same $\gamma$ as for the time 
scale. This value of $\phi_p^G$ is too far from that obtained using $\tau_q$. 
Therefore, we cannot have similar $\phi_p^G$ and $\gamma$ to explain the 
behaviour of both $\tau_q$ and $D_s$, implying that the MCT prediction, eq. 
(\ref{power-law}) is violated.

\begin{figure}
\psfig{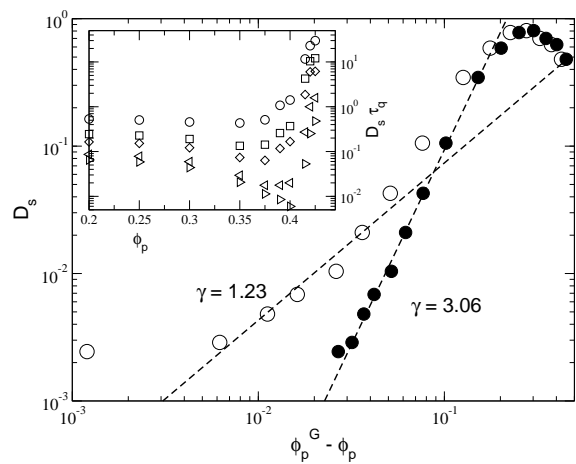}
\caption {\label{ds}
Self diffusion coefficient, $D_s$, vs. $\phi_p^G-\phi_p$ for two values of
$\phi_p^G$: $\phi_p^G=0.4519$ (closed circles) and $\phi_p^G=0.4265$ (open
circles). The dashed lines are the power law fittings to the data, with the 
exponent shown in the figure. Inset: $D_s\tau_q$ vs. polymer fraction for
different wave-vectors. Symbols as in Figure \ref{tauq_phi}.}
\end{figure}

In Figure \ref{ds}, we present $D_s$ vs. $\phi_p^G-\phi_p$ using for
$\phi_p^G$ both the value estimated from $\tau_q$ and that from $D_s$. We
consider more desirable to have similar $\phi_p^G$ to explain the behaviour
of $\tau_q$ and $D_s$, even though this implies two different $\gamma$:
$\gamma=3.1$ for $\tau_q$ and $\gamma=1.23$ for $D_s$. As obtained in other
non-ergodicity transitions \cite{kob95,Aichele01,horbach01}, the $\gamma$
exponent is lower in the diffusion coefficient than in the time scale,
although the difference between both values of $\gamma$ is bigger in our case.

In order to stress the different $\gamma$ exponents in the divergence of the
time scale and $1/D_s$, we have plotted $D_s \tau_q$ as a function of
$\phi_p$ for different wave-vectors in the inset to figure \ref{ds}. This
product, that should be constant according to MCT, diverges as the polymer
fraction approaches $\phi_p^G$. The divergence follows a power law with the
exponent equal the difference between both values of $\gamma$.

The maximum in the self diffusion coefficient (upper-right corner of Figure
\ref{ds}) is a consequence of the re-entrant glass transition at high packing
fractions \cite{pham02,foffi02a}. A weak short range attraction at first 
destabilizes the cage and thus the glass transition moves to higher particle 
concentration initially as the polymer fraction is increased. At constant 
colloid concentration the diffusion thus first speeds up with increasing 
$\phi_p$, until for intermediate attraction strengths the gel line is 
approached, where the opposite trend then dominates. At $\phi_c=0.40$, the 
glass transition is rather far removed and thus has little effect, but the 
increase of $D$ is still measurable and the diffusion coefficient can be used 
as a measure of the distance to the closest transition. The maximum thus 
indicates the re-entrant shape of the non-ergodicity line. 

\begin{figure}
\psfig{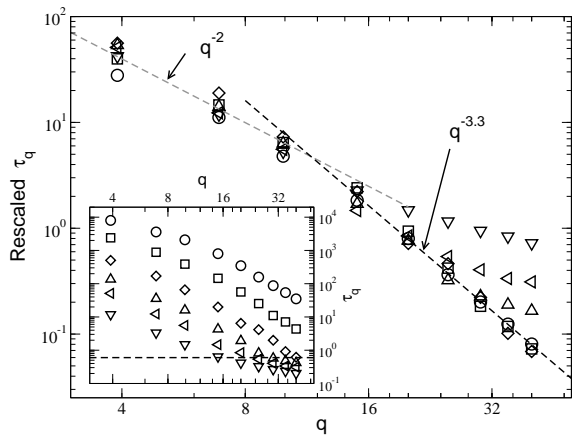}
\caption {\label{tauq_q}
Inset: Time scale, $\tau_q$ as a function of the wave-vector, $q$, for 
different states close to the gel: $\phi_p=0.42$ (circles), $\phi_p=0.415$ 
(squares), $\phi_p=0.41$ (diamonds), $\phi_p=0.40$ (upward triangles), 
$\phi_p=0.39$ (left-ward triangles) and $\phi_p=0.375$ (downward triangles). 
Main body: Same data, re-scaled to collapse in the low-q
power-law behavior. 
The dashed lines show power-law behaviours with
exponents $2$  (gray line) and $3.3$ (black line).}
\end{figure}

The wave-vector dependence of the time scale $\tau_q$ can also be compared 
with theoretical predictions. At low $q$, the time scale is expected to 
behave as $q^{-2}$, corresponding to a diffusive process over large distances. 
Yet, because the simulated scattering functions exhibit
non--exponential relaxation even for the smallest wavevectors, this
simple theoretical scenario is not expected to appear in our case. At 
intermediate wave-vectors, where the Kohlrausch exponent becomes
comparable to the von Schweidler one the theory predicts a 
decrease as $q^{-1/b}$, whereas at even higher $q$, the distances involved are 
dominated by the microscopic dynamics, and corrections to this behaviour are 
expected \cite{fuchs99}. The inset to Figure \ref{tauq_q} shows $\tau_q$ for 
different states close to the gel transition. In order to make clear common 
properties the curves have been scaled vertically to collapse (main figure). 

It can be seen in this figure that the behaviour of $\tau_q$ at low 
wave-vectors (below $q=10$), indeed shows a $q^{-2}$ behavior, which however
is not the one explained by MCT. 
At higher $q$, another power-law trend is observed, with a higher exponent: 
$q^{-3.3}$. The crossover from the low-$q$ behaviour to the high-$q$ one,
compares nicely with the wave-vector where the Kohlraush exponent becomes 
equal to the von Schweidler one (Figure \ref{a_q}). The exponent of the 
high-$q$ region yields $b=0.30$, lower than the value obtained 
from the analysis in Figure \ref{fsqt}. However, this value is 
quite close to the measured von Schweidler value and much smaller 
than the HSS one. Deviations from this power-law behaviour are observed at 
high $q$ for the lowest $\phi_p$ presented in the figure. These deviations 
are caused by the microscopic dynamics, as they occur when $\tau_q$ is lower 
than a certain value, regardless the polymer fraction. This value, presented 
in the inset as an horizontal line, is $t_0\sim 0.6$, which agrees with the 
time one would estimate from the correlators in Figure \ref{fsqt}.

\subsection{Mean Squared Displacement}

\begin{figure}
\psfig{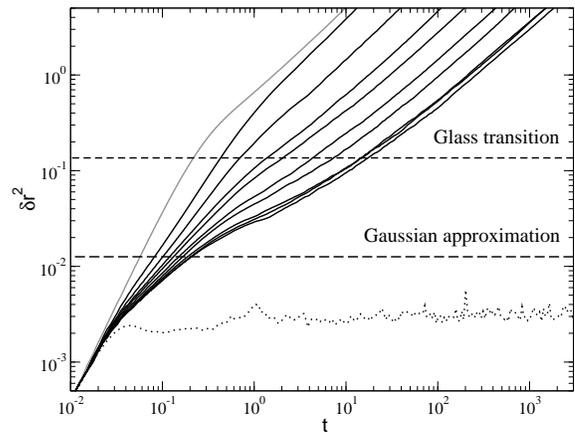}
\caption {\label{msd}
Mean squared displacement of the particles as the gel transition is approached.
Black curves from left to right: $\phi_p=0.30, 0.325, 0.35, 0.375, 0.39, 0.40,
0.41, 0.415, 0.42, 0.425$. Grey curve: $\phi_p=0$. The dashed horizontal 
lines indicate the localization length corresponding to the glass transition
(short dash) in the HSS and the estimate from Figure \ref{fq} (long dash). 
Dotted curve: mean squared displacement for a single particle in a frozen 
environment at $\phi_p=0.425$.}
\end{figure}

We turn now our attention to the MSD curves, 
that were partially analysed to obtain the diffusion coefficients presented 
in Figure \ref{ds}. We are only interested in the slowing down close to the
gel transition and thus we do not show the MSD for low polymer fractions,
where the attraction speeds up the dynamics and increases the diffusivity
(see figure 9 and \cite{pham02}). The MSD, after a short initial regime of
free flight, $\delta r^2 \propto t^2$, slows down because of the particle
interactions and takes longer and longer to reach the long-time regime
diffusive, where $\delta r^2=6D_st$. An important feature that can be 
obtained from the MSD of the particles in the system, is the localization
length, where the particle interactions hinder particle motion most strongly, 
and in the idealized glass state, arrest it. It can be compared with the 
estimate using the Gaussian approximation (see Fig. \ref{fq}). In Figure 
\ref{msd} we present the MSD for increasing polymer volume fractions. As the 
gel transition is approached, the localization length shows up as an 
indication of a plateau, signaling the bond formation. As already discussed 
above, $r_l$ is much shorter than in the HSS glass transition (upper dashed line 
in Figure \ref{msd}), because of the driving mechanism. 

The lower dashed line in this figure is the localization length, as estimated 
from the non-ergodicity parameter using the Gaussian approximation 
($r_l^2=0.0126$). Although a clear plateau has not fully developed in our 
curves, its height seems to be above that estimate, by a factor 
$\sim 1.5-2$. Since the Gaussian approximation works very well in the case 
of the HSS, this suggests big non-Gaussian corrections at the gel transition. 
Before testing the Gaussian approximation, we stress that the localization
length gives a typical size of the mesh of bonds formed between neighbouring 
particles, and that the slow structral units are continuously and 
cooperatively rearranging. In order to test this idea about a correlated 
region which cooperatively rearranges with and around each particle, a single
mobile particle is considered in a fixed environment. A well equilibrated 
system with $\phi_p=0.425$ is frozen, and  only one particle is allowed to 
move. This mobile particle now explores a {\sl frozen} environment, 
providing the {\sl structural} size of the region it is confined to. The mean 
squared displacement so obtained is given in Figure \ref{msd} (dotted line).
Some particles (1.6 \%) were able to break their bonds and diffuse freely in 
the frozen environment. For the particles that stay localized, it can be 
observed that the length of the frozen bonds is much smaller than the 
localization length. This fact demonstrates that the structure of bonds, 
like the repulsive cage at the glass transition in the HSS or LJS, 
is dynamic, and constantly rearranges cooperatively. This collective 
restructuring of the system fluidizes it, and restores ergodicity, which 
cannot occur in the frozen system, where the particles are not able to 
diffuse even at very long times.

\begin{figure}
\psfig{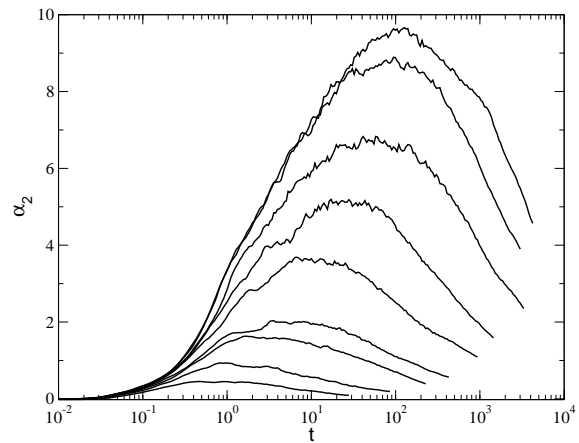}
\caption {\label{alpha_2}
Non-Gaussian parameter $\alpha_2$ as a function of time for states approaching
the gel transition at the same states as Figure \ref{msd}. The maximum
increases with increasing $\phi_p$.}
\end{figure}

We turn now back to the Gaussian approximation, and its accuracy. Usually, 
this is tested by measuring the non-Gaussian parameter, defined as:

\begin{equation}
\alpha_2\:=\:\frac{3 \langle r^4 (t) \rangle}{5 \langle r^2(t) 
\rangle^2} \,-\,1
\end{equation}

\noindent where the averages imply ensemble averaging.
This parameter measures the deviation of the probability density 
function for the single particle motion from Gaussian behaviour, and vanishes
for diffusive motion. Special care must be taken when performing the ensemble
averages in polydisperse systems, as pointed out in \cite{doliwa99}. The
non-Gaussian parameter must be calculated for every particle (the averages in
the definition above thus implying time-origin averaging only), and particle
averaging is taken on the values of $\alpha_2$ (so long as long enough time 
intervals are studied, each particle will sample the distribution relevant to 
its own size in an ergodic fashion). The non-Gaussian parameters 
for states with increasing $\phi_p$ are presented as a function of time in 
Figure \ref{alpha_2}. At short times $\alpha_2$ tends to zero, since the 
system shows Gaussian behaviour during its unhindered ballistic regime. At 
long times, when the particles break free from their bonds and hydrodynamic 
diffusion holds, $\alpha_2$ again  goes back towards zero. At intermediate 
times, corresponding to the plateaus in both the correlation function and the 
mean squared displacement, $\alpha_2$ grows, since the single particle
motion hindered by bonding is not Brownian. As a result, $\alpha_2$ shows 
a maximum, whose height and position grows in time, because the particles take
longer and longer to break free and start diffusing.

The behaviour of the simulated $\alpha_2$ obeys the general expectations
\cite{donati99,donati99a,doliwa00}, but 
important differences are observed in the comparison with the results for the 
HSS or LJS. Whereas in those cases the height of the maximum for similar (or 
even higher) $\alpha$-relaxation times is around $2$, at the gel transition 
much higher values are measured. Another interesting difference is the failure
of the short time scaling, observed both in the HSS and LJS. Both effects can 
be rationalized considering that the {\sl cage} is indeed a network of bonds 
in the case of a gel, rather than a cavity. The strength of these bonds is given
by the intensity of the interaction, and thus, it is modified for different 
states, disabling the short time collapse. Because the bonds are short ranged,
they affect the particle motion from very short times onward, so that the 
particles {\sl feel} the hindrance much longer in the gel case. 

It can be concluded that the non-Gaussian corrections are very important in
the gel transition. Therefore, the localization length estimated from the 
non-ergodicity parameter may be inaccurate, as discussed above. However, it 
still provides an indication of how small the localization length is.
A better indication of $r_l$ can be obtained within the Gaussian approximation
if only low wave-vectors are used in fitting expression 
(\ref{gauss}). The fitted curve is presented in Figure \ref{fq} by the dashed 
line, where only the three lowest $q$'s are fitted. The estimated $f_q^s$ 
deviates from the data at higher wave-vectors, showing high 
non-Gaussian corrections. The localization length is higher than the previous
value: $r_l^2=0.0162$. Thus, this fitting provides data more consistent with 
the MSD curves and the non-Gaussian parameter.

\subsection{Higher Colloid Volume Fraction}

We move now to a higher colloid volume fraction: $\phi_c=0.50$. These
results are presented to supplement the findings at the lower packing fraction
and test for the prediction of stronger stretching closer to the higher
order singularity. As indicated 
in the theoretical section, MCT predicts a higher order singularity in the 
vicinity of the junction of the gel and glass lines; i.e. at high polymer and 
colloid densities. In this particular system we found clear indications of 
this singularity in simulations at $\phi_c=0.55$ and $\phi_p=0.375$ 
\cite{puertas02}. The isochore under study now, $\phi_c=0.50$, could be close 
enough to the higher order singularity to show some effects.

\begin{figure}
\psfig{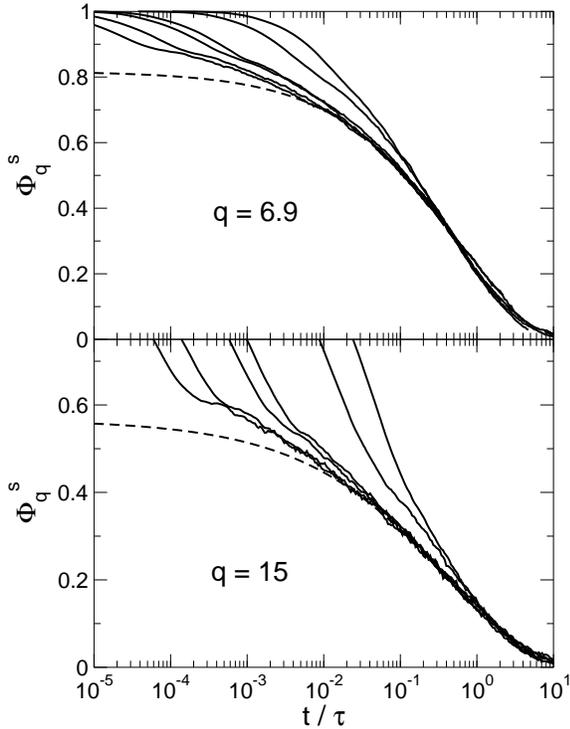}
\caption {\label{scaling50}
Correlation functions for $\phi_c=0.50$ and different polymer concentrations. 
>From left to right: $\phi_p=0.35,\,0.36,\,0.375,\,0.38,\,0.385,\,0.39$. Two 
wave-vectors are studied, as labeled in the figures. The dashed lines 
represent the KWW fittings to the $\phi_c=0.40$ correlation functions.}
\end{figure}

In Figure \ref{scaling50} we present the correlation functions for increasing 
polymer fractions at the same wave-vectors as Figure \ref{scaling}, re-scaled 
to collapse in the long-time decay. It is interesting to note that the polymer 
concentrations studied in this case are lower than those studied at the lower
colloid volume fraction. In accordance with  experiments and theory, this 
indicates that the gel transition takes place at lower polymer fractions the 
higher the colloid concentration. 

In Figure \ref{scaling50}, it can be observed that the correlators do not 
collapse over the whole $\alpha$-decay, but only in the end. These deviations 
are expected because of the higher order singularity, which 
is at higher densities. However, we stress that although this singularity has
clear effects on the correlation functions, they do not show so clear 
signatures as that of the $\phi_c=0.55$ isochore \cite{puertas02}. At this high
concentration, a logarithmic decay was observed, with a wave-vector dependent
extension.

Because these corrections affect the early $\alpha$-decay, analyzing the 
correlation functions is difficult. Furthermore, the plateau is not observed, 
and the von Schweidler analysis is thus extremely difficult. In order to 
analyse the self intermediate scattering function, we compare the stretching 
of the curves at $\phi_c=0.40$ and $\phi_c=0.50$; in Figure \ref{scaling50} 
the $\alpha$-decay master function of the $\phi_c=0.40$ state, as parametrized
by the KWW fitting is included. It can be seen that this curve can be rescaled
to collapse onto the $\alpha$-decay of the correlators at $\phi_c=0.50$ for both 
wave-vectors at long times. This indicates that the von Schweidler 
exponent is very similar in both cases, but also points out the effect of
the high order singularity. According to MCT, $b$ should decrease as the
singularity is approached, but this behaviour is not observed in our case.
Comparison of the $\alpha$-decays by fitting the 
KWW stretched exponential to the master function is troublesome, since only 
the late decay is obtained unambiguosly.

\begin{figure}
\psfig{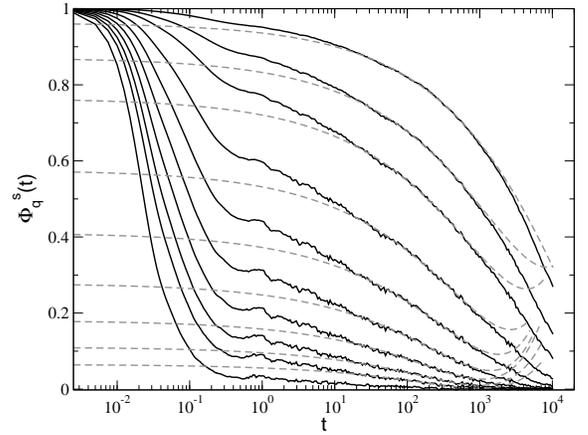}
\caption {\label{fsqt50} Intermediate scattering function (self part), 
$\Phi_q^s$, for the same wave-vectors as Figure \ref{fsqt} wave-vectors for
$\phi_c=0.50$ and $\phi_p=0.39$. The dashed lines are fittings from eq.
(\ref{a-decay}) up to second order, with the same von Schweidler exponent
as $\phi_c=0.40$.}
\end{figure}

The similarity of both $\alpha$-decays was used in the von Schweidler analysis
of the correlation function, and only the non-ergodicity parameter and 
amplitudes were fitted. Since the upper part of the decay is known to be 
affected by the higher order singularity close-by, that part must be discarded
in the fittings. The correlation functions and fittings are presented in 
Figure \ref{fsqt50} for the state $\phi_c=0.50$ and $\phi_p=0.39$, for the
same wave-vectors as Figure \ref{fsqt}. The main conclusion is that the late
$\alpha$-decay at all wave-vectors can be correctly described by the von 
Schweidler decay, with the same exponent as the state at $\phi_c=0.40$. 
The non-ergodicity parameters obtained 
from the fitting are slightly lower than those of $\phi_c=0.40$, but similar
within the error bars. According to MCT, $f_q^s$ decreases when approaching the
glass part of the non-ergodicity line (signaling an increase in the 
localization length). Our result is thus consistent with this prediction. 

\begin{figure}
\psfig{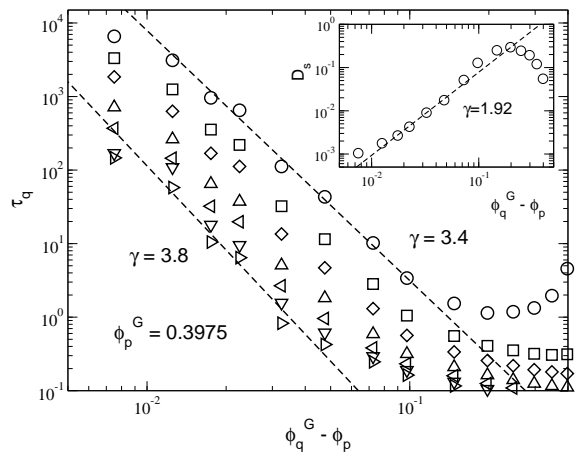}
\caption {\label{tauq_phi50}
Wave-vector dependent time scale, $\tau_q$, vs. $\phi_p^G-\phi_p$ for
different wave-vectors; symbols from top to botton correspond to 
$q=3.9,\,6.9,\,9.9,\,15,\,20,\,25,\,30$. The lines are power law fittings to 
$q=3.9$ and $q=30$. $\phi_p^G$ and $\gamma$ were fitted.}
\end{figure}

With these values of the non-ergodicity parameter one can define also the 
wave-vector dependent time scale, $\tau_q$, as discussed above. In order to 
test the value of the von Schweidler exponent, using
eq. (\ref{power-law}) we have performed a three parameter fitting to obtain
$\gamma$ and $\phi_p^G$. In Figure \ref{tauq_phi50}, $\tau_q$ is presented
as a function of $\phi_p^G-\phi_p$ for different wave-vectors. The power-law
fittings for two wave-vectors are also plotted and the critical polymer 
fraction $\phi_p^G$ is given.  

The values of $\gamma$ obtained from this analysis for different wave-vector 
range from $\gamma=3.37$ to $\gamma=3.82$, the mean value being $\gamma=3.70$. 
This value of $\gamma$ implies a smaller von Schweidler exponent, $b=0.33$, in 
disagreement with our previous estimate, but backing the MCT prediction.
Using the same value of $\phi_p^G$, the vanishing of the  self-diffusion 
coefficient, $D_s$, can be analysed, and is presented in the inset to this 
figure. A power-law is observed in this case, with an exponent, $\gamma=1.92$,
which, again in contradiction to MCT, leaves us with a big difference between 
the two values of $\gamma$.

The diffusion coefficients in the inset of figure \ref{tauq_phi50} again 
indicate the re-entrat glass transition. They describe a maximum, more 
pronounced than that observed in Figure \ref{ds} because the glass line is 
closer to the $\phi_c=0.50$ isochore. The minimum in $\tau_q$, which is 
observed only for $q=3.9$, in an equivalent way indicates the  shape of the 
non-ergodicity transition line. At higher wave-vectors, the glass transition 
causes very low $f_q^s$ and the time scales merge with the microscopic 
transient and thus this feature is suppressed.

The wave-vector dependence of $\tau_q$ can also be studied, as done for the 
lower concentration, yielding another estimate of $b$. In this case, a 
similar plot as Figure \ref{tauq_q} is obtained, where the low $q$ region is 
compatible with a $q^{-2}$ behaviour, and a higher exponent at higher $q$, 
yielding a value of $b=0.38$. This value is in agreement with the nice 
comparison between the $\phi_c=0.40$ and $\phi_c=0.50$ isochores, but not 
with $\gamma$ or the MCT prediction. We may then conclude that analysis of 
this state is extremely difficult, but our indications state that the von 
Schweidler exponent is similar for both packing fractions, but probably 
slightly lower in the higher concentration.

Finally, we would like to point out that the non-Gaussian parameter at this 
packing fraction shows a behaviour similar to that shown in Figure 
\ref{alpha_2}, i.e. the peak is as high, and no short-time scaling is observed.

\section{Conclusions}

In this paper, by means of simulations, we have tested the universal 
predictions of MCT for gelation in colloidal systems, viewed as an 
attraction--driven glass transition. The 
self parts of the intermediate scattering function for states close to this 
transition have been analysed and the results were compared with the 
theoretical predictions. For the $\phi_c=0.40$ isochore, which is far enough
from the high order singularity, the correlation functions can be 
$\alpha$-scaled. The time scale of the $\alpha$-decay was shown to obey a 
power law divergence, with an exponent, $\gamma$, related to the von Schweidler
exponent, obtained from the early $\alpha$-decay. Both features are predicted 
by MCT for all non-ergodicity transitions. Also, the wave-vector analysis of 
the time scale follows the behaviours predicted by MCT, with a small 
difference in the value of the von Schweidler exponent. 

The wave-vector analysis of the correlation functions depends on details of 
the interaction potential, and thus provides information about the mechanism, 
leading to the transition. In our case, it establishes that the gel transition is 
driven by a short-range mechanism, namely, bond formation, as observed
in the pair distribution function. Additionally, it has been shown 
that the KWW stretched exponential can account for the $\alpha$-decay of the 
correlation functions, as in other non-ergodicity transitions.

We have also tested the Gaussian approximation, which works very well for 
the HSS. The non-Gaussian parameter, $\alpha_2$, establishes that this 
approximation is much worse in the case of the gel transition than for the 
glass transition. It was also tested when comparing the estimated localization
length from the non-ergodicity parameter with the MSD of the particles. The 
diffusion coefficient has been also studied. It tends to zero as the 
transition is approached following a power-law, with an exponent much lower 
than $\gamma$, in accordance with simulations of glass transitions in other 
systems, but in disagreement with MCT, where both exponents are equal. 

Finally, when the colloid concentration is increased, the system shows 
signatures of the high order singularity nearby and little can be discussed
about the exponents $b$ or $\gamma$. However, only slight changes in the
numbers are expected, since the qualitative behaviour is reproduced, except
for the $\alpha$-scaling. Also, the diffusion coefficient follows a 
power law with a different exponent and the non-Gaussian parameter reaches 
values similar to the $\phi_c=0.40$ case.

Therefore, our main conclusion is that MCT accounts for most features of the 
simulated systems on approach to the gel transition, but the discrepancies 
already found in other non-ergodicity transitions (such as the repulsion-driven 
glass transition in hard sphere systems) are also obtained here.

\begin{center}
{\sc Acknowledgements}
\end{center}

The authors thank W. Kob and R. Sear for useful discussion. A.M.P. 
acknowledges the financial support by the CICYT (project MAT2000-1550-CO3-02).
M.F. was supported by the DFG under grant Fu 309/3.


\begin{thebibliography}{99}
\bibitem{note1} This state was carefully checked for aging or
out-of-equilibrim effects, but within our tests, it is fully equilibrated.
\end{thebibliography}
\end{document}